\documentclass[prd,amsmath,amssymb,showpacs,superscriptaddress,nofootinbib,twocolumn]{revtex4-1}
\usepackage{graphicx}
\usepackage{epsfig,graphics,subfigure,psfrag,amsmath,amssymb}
\usepackage{lineno}
\usepackage{dcolumn}
\usepackage{bm}
\usepackage{overpic}
\usepackage[english]{babel}
\usepackage{color}
\usepackage{multirow}
\usepackage{colortbl}
\usepackage{epstopdf}
\usepackage[colorlinks,linkcolor=black,anchorcolor=blue,citecolor=blue]{hyperref}
\usepackage{gensymb}
\usepackage{ulem}

\newcommand{\BR}{{\cal B}}

\newcommand{\jpsi}{J/\psi}

\usepackage[
total={6.5in,8.75in}, top=1.2in, left=0.9in, 
]{geometry}

\begin{document}

\title{\texorpdfstring{\boldmath
Search for $\psi(3686)\to \gamma\eta_c(\eta(1405))\to \gamma \pi^{+}\pi^{-}\pi^{0}$ } {search for }}


\author{M.~Ablikim$^{1}$, M.~N.~Achasov$^{9,d}$, S. ~Ahmed$^{14}$, M.~Albrecht$^{4}$, A.~Amoroso$^{50A,50C}$, F.~F.~An$^{1}$, Q.~An$^{47,39}$, J.~Z.~Bai$^{1}$, O.~Bakina$^{24}$, R.~Baldini Ferroli$^{20A}$
% ...
\\
 \vspace{0.2cm}
 (BESIII Collaboration)\\
 \vspace{0.2cm} {\it
$^{1}$ Institute of High Energy Physics, Beijing 100049, People's Republic of China\\
% ...
$^{4}$ Bochum Ruhr-University, D-44780 Bochum, Germany\\
% ...
$^{9}$ G.I. Budker Institute of Nuclear Physics SB RAS (BINP), Novosibirsk 630090, Russia\\
% ...
$^{14}$ Helmholtz Institute Mainz, Johann-Joachim-Becher-Weg 45, D-55099 Mainz, Germany\\
% ...
$^{20}$ (A)INFN Laboratori Nazionali di Frascati, I-00044, Frascati, Italy; (B)INFN and University of Perugia, I-06100, Perugia, Italy\\
% ...
$^{24}$ Joint Institute for Nuclear Research, 141980 Dubna, Moscow region, Russia\\
% ...
$^{39}$ State Key Laboratory of Particle Detection and Electronics, Beijing 100049, Hefei 230026, People's Republic of China\\
% ...
$^{47}$ University of Science and Technology of China, Hefei 230026, People's Republic of China\\
% ...
$^{50}$ (A)University of Turin, I-10125, Turin, Italy; (B)University of Eastern Piedmont, I-15121, Alessandria, Italy; (C)INFN, I-10125, Turin, Italy\\
% ...
$^{d}$ Also at the Novosibirsk State University, Novosibirsk, 630090, Russia\\
% ...
}
}

\begin{abstract}
  Using a sample of $448.1\times10^{6}$ $\psi(3686)$ events collected
  with the BESIII detector, a search for the isospin violating decay
  $\eta_{c}\to\pi^{+}\pi^{-}\pi^{0}$ via $\psi(3686)\to\gamma\eta_{c}$ is presented. No signal is observed, and the
  upper limit on $\mathcal{B}(\psi(3686)\to \gamma\eta_{c})\times\mathcal{B}(\eta_{c}\to
  \pi^{+}\pi^{-}\pi^{0} )$ is determined to be $1.6\times10^{-6}$ at
  the $90\%$ confidence level.  In addition, a search for
  $\eta(1405)\to f_{0}(980)\pi^{0}$ in $\psi(3686)$ radiative decays
  is performed.  No signal is observed, and the branching
  fraction $\mathcal{B}(\psi(3686)\to\gamma\eta(1405))\times \mathcal{B}(\eta(1405)\to f_{0}(980)\pi^{0})\times \mathcal{B}(f_{0}(980)\to\pi^+\pi^- )$ is calculated to be
  less than $ 5.0\times10^{-7}$ at the $90\%$ confidence level.
\end{abstract}

\pacs{ 13.25.Gv, 14.40.Be, 12.38.Qk, 11.30.Er }
\maketitle

\section{Introduction}
As the lowest-lying  $c\bar{c}$ state, the pseudoscalar meson $\eta_c$
has attracted considerable theoretical and experimental attention
since it was discovered three decades ago~\cite{etac}. 
 To the lowest order in perturbation theory, the $\eta_c$ decays through $c \bar{c}$
annihilation into two gluons.  The $\eta_c$ is then expected to have
numerous hadronic decay modes into two- or three-body hadronic final states,
and many of them have been measured~\cite{PDG2016}.  However, the
three-pion decay mode has not yet been studied, but its measurement is important to test isospin symmetry~\cite{isospin1,isospin2,isospin3}. 

 Charmonium radiative decays, especially those of
$J/\psi$ and $\psi(3686)$, provide an excellent laboratory  for the study of neutral pseudoscalar  meson decays. For example, the BESIII
experiment using $J/\psi$
radiative decays has performed a series of
analyses on three pion decays~\cite{beseta1,beseta2,beseta3,beseta4,beseta5,beseta6}, and for the first time reported the observation of the isospin violating decay  $\eta(1405)\rightarrow 3\pi$~\cite{wuzhi}. Of
particular interest in $\eta(1405)\rightarrow 3\pi$ decay  is a narrow
structure around 0.98 GeV/c$^2$ in the $\pi\pi$ mass spectrum, identified with the $f_0(980)$, which can be  interpreted  under the  triangle
singularity scheme~\cite{zhaoqiang,zhaoqiang1,zhaoqiang2}.

In this analysis, we
perform a search for the isospin violating decay
$\eta_c\to\pi^+\pi^-\pi^0$ using a sample of
$448.1\times10^{6}$ $\psi(3686)$ events~\cite{psi'data} collected with
the BESIII~\cite{Ablikim2010345} detector operating at the BEPCII~\cite{BEPCII} storage ring. We also perform a search for $\eta(1405)\to f_0(980)\pi^0$ in the $\psi(3686)$ radiative decays to test the
``12\% rule''~\cite{12rule1,12rule2,12rule3}, in which perturbative QCD predicts  the ratio of the branching fractions of $\psi(3686)$
and $J/\psi$ into the same final hadronic state is given

\begin{equation}
      \begin{aligned}
      Q = \frac{\mathcal{B}_{\psi(3686)\to h}}{\mathcal{B}_{J/\psi\to h}}=\frac{\mathcal{B}_{\psi(3686)\to l^{+}l^{-}}}{\mathcal{B}_{J/\psi\to l^{+}l^{-}}}\approx (12.4\pm0.4)\%.
\end{aligned}
\end{equation}
The rule is expected to also hold for radiative decays
to the same final hadronic state.

\section{\texorpdfstring{Detector and Monte Carlo Simulation}{Detector and MC Simulation}}
BEPCII is a double-ring $e^{+}e^{-}$ collider providing a peak
luminosity of $10^{33}$ cm$^{-2}$s$^{-1}$ at a beam energy of 1.89
GeV.  The BES\uppercase\expandafter{\romannumeral3}
detector~\cite{Ablikim2010345} consists of a helium-based main drift
chamber (MDC), a plastic scintillator time-of-flight (TOF) system, a
CsI(Tl) electromagnetic calorimeter (EMC), and a multi-layer resistive
plate chamber muon counter system. With a geometrical acceptance of
93\% of 4$\pi$, the BESIII detector operates in a magnetic field of
1.0~T provided by a superconducting solenoidal magnet.

Monte Carlo (MC) simulations are used
to determine detector efficiency, optimize event selection and
estimate backgrounds. The BES\uppercase\expandafter{\romannumeral3}
detector is modeled with GEANT4 ~\cite{GEANT}.  For the inclusive MC,
the production of the $\psi(3686)$ resonance is simulated by the MC
event generator KKMC~\cite{ref:kkmc,ref:kkmc2}, and the decays are
generated by EVTGEN~\cite{ref:evtgen,ref:evtgen2} for known decay modes with
branching fractions being set to Particle Data Group (PDG)~\cite{PDG2016}  world average values,
 while the remaining unknown decays are generated by
LUNDCHARM~\cite{ref:lundcharm}.  For
$\psi(3686)\rightarrow\gamma\eta_{c},~\eta_{c}\rightarrow\pi^{+}\pi^{-}\pi^{0}$
decays, the line shape of the $\eta_{c}$ meson is described by
$E_{\gamma}^{7}\times {|BW(m)|}^2\times D(E_{\gamma})$, where $m$ is the
$\pi^{+}\pi^{-}\pi^{0}$ invariant mass,
$E_{\gamma}=\frac{M_{\psi(3686)}^{2}-m^2}{2M_{\psi(3686)}}$ is the
energy of the transition photon in the rest frame of $\psi(3686)$,
 $BW(m)=\frac{1}{m^2-M_{\eta_c}^2+iM_{\eta_c}\Gamma_{\eta_c}}$ is a relativistic Breit-Wigner function,
 $M_{\eta_c}$ and $\Gamma_{\eta_c}$ are the mass and width of $\eta_c$,
 $D(E_{\gamma})=\frac{E_{0}^{2}}{E_{0}E_{\gamma}+(E_{0}-E_{\gamma})^2}$
is a function introduced by the KEDR collaboration~\cite{KEDR}, which
damps the low-mass divergent tail, where
$E_{0}=\frac{M_{\psi(3686)}^{2}-{M_{\eta_c}}^2}{2M_{\psi(3686)}}$ is
the peak energy of the transition photon.  In the MC simulation,
$\eta_{c}\rightarrow\pi^{+}\pi^{-}\pi^{0}$ events are generated
according to a phase space distribution.

\section{Data analysis}
\subsection{\texorpdfstring{\boldmath  $\psi(3686)\to\gamma\eta_c,\eta_{c}\to \pi^{+}\pi^{-}\pi^{0}$}{etac to gamma3pi}}
For $\psi(3686)\rightarrow \gamma\eta_c$ with $\eta_c$ subsequently
decaying into $\pi^+\pi^-\pi^0$, the final state in this analysis is
$\pi^+\pi^-\gamma\gamma\gamma$.  Charged tracks must be in the active
region of the MDC, corresponding to $|\cos\theta|<$ 0.93, where
$\theta$ is the polar angle of the charged track with respect to the
beam direction, and are required to pass within $\pm$10 cm of the
interaction point in the beam direction and 1 cm of the beam line in
the plane perpendicular to the beam.  Photon candidates must have
minimum energies of 25 MeV in the EMC barrel ($|\cos\theta|<0.8$) or
50 MeV in the EMC end-caps ($0.86<|\cos\theta|<0.92$). To eliminate
photons radiated from charged particles, each photon must be separated by at
least $10\degree$ from any charged track.  A requirement on the photon
time, $TDC$, in the EMC, $0\leq TDC \leq14$ (50 ns/count), is used to
suppress noise and energy deposits unrelated to the event. Events with
two oppositely charged tracks and at least three photons are selected
for further analysis.  The two charged tracks are required to be
identified as pions using the combined information of dE/dx from the
MDC and the flight time from the TOF.

A four-constraint (4C) kinematic fit imposing energy-momentum
conservation is performed under the $\gamma\gamma\gamma\pi^{+}\pi^{-}$
hypothesis, and the fit results are used for the kinematic quantities
below.  If there are more than three photon candidates in an event,
the combination with the smallest $\chi^{2}_{4C}$ is retained, and
$\chi^{2}_{4C}$ is required to be less than 20.  To suppress the
background events with two or four photons in the final states, 4C
kinematic fits are also performed under the $\gamma\gamma \pi^+\pi^-$
and $\gamma\gamma\gamma\gamma \pi^+\pi^-$ hypotheses, and
$\chi^{2}_{4C}$ is required to be less than the $\chi^2$ values of the
$\gamma\gamma\pi^{+}\pi^{-}$ and
$\gamma\gamma\gamma\gamma\pi^{+}\pi^{-}$ hypotheses.  To select
$\pi^0$ candidates, the invariant mass of two photons,
$M_{\gamma\gamma}$, must satisfy
$|M_{\gamma\gamma}-m_{\pi^{0}}|<0.015$ GeV/$c^{2}$, where
$m_{\pi^{0}}$ is the nominal $\pi^0$ mass~\cite{PDG2016}. If more than
one $\gamma\gamma$ combination satisfies this requirement, the one
with $M_{\gamma\gamma}$ closest to $m_{\pi^{0}}$ is selected.
To reject background events with an $\eta$ in the final state, we
require that the invariant masses of the other two possible photon pairs
are not within the $\eta$ mass region, $|M_{\gamma\gamma}-m_{\eta}|>0.02$
GeV/$ c^{2}$, where $m_{\eta}$ is the nominal $\eta$ mass~\cite{PDG2016}. In order to
reduce the $\omega\rightarrow\gamma\pi^0$ background,
$|M_{\gamma\pi^{0}}-m_{\omega}|>0.05 $ GeV/$c^{2}$ is required, where
$M_{\gamma\pi^{0}}$ and $m_{\omega}$ are the $\gamma \pi^0$
invariant mass and nominal $\omega$ mass~\cite{PDG2016}, respectively. Events with
a $\gamma \pi^+ \pi^-$ invariant mass in the vicinity of the
$J/\psi$ ($|M_{\gamma\pi^{+}\pi^{-}}-m_{J/\psi}|< 0.02$ GeV/$c^{2}$) are vetoed to suppress
background events from $ \psi(3686)\to \pi^{0}J/\psi$
($J/\psi\to\gamma\pi^{+}\pi^{-} $ or $
J/\psi\to\pi^{+}\pi^{-}\pi^{0}$  with a missing photon from the $\pi^0$).

\begin{figure}[htbp]
    \centering
    \vskip -0.0cm
    \hskip -0.0cm \mbox{
    \begin{overpic}[width=7.5cm,height=5.0cm,angle=0]{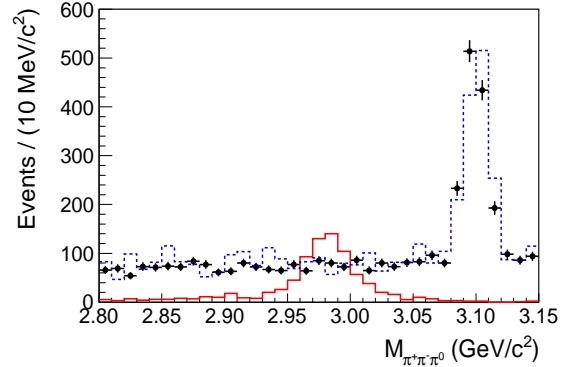}
    \end{overpic}} {\caption{ The distributions of
      $M_{\pi^{+}\pi^{-}\pi^{0}}$ in the vicinity of the
      $\eta_{c}$. Dots with error bars are data, the solid line
      histogram is the $\eta_c$ line shape from the exclusive MC
      simulation, and the dashed line  are the backgrounds estimated from  inclusive MC sample and  initial state radiation process $e^+e^-\to \gamma_\text{ISR}J/\psi$.
    }
    \label{pippimpi0}}
\end{figure}

After the above requirements, the $M_{\pi^{+}\pi^{-}\pi^{0}}$
distribution is shown in Fig.~\ref{pippimpi0}, where no clear $\eta_c$
signal is seen. Possible backgrounds are studied with an inclusive MC
sample of $5.06\times 10^8$ $\psi(3686)$ decays, and the background
events contributing to the $J/\psi$ peak in Fig.~\ref{pippimpi0} are
dominantly from $\psi(3686)\rightarrow\pi^0 J/\psi,\,
J/\psi\to\pi^+\pi^-\pi^0$ and $\psi(3686)\to\gamma\chi_{cJ},\,
\chi_{cJ}\rightarrow\gamma J/\psi,\, J/\psi\rightarrow
\pi^+\pi^-\pi^0$, while the other background events, mainly from
$\psi(3686)\rightarrow\rho\pi\pi$, contribute a smooth shape in the
$\eta_c$ mass region. Using the off-resonance continuum data sample
taken at a center-of-mass energy of 3.65 GeV, corresponding an
integrated luminosity of 44 pb$^{-1}$~\cite{lu}, we also investigate the
background events from QED processes. There are no peaking
contributions except for a small  $J/\psi$ peak due
to the initial state radiation process $e^+e^-\to\gamma_\text{ISR}J/\psi$.

\begin{figure}[htbp]
    \centering
    \vskip 0.0cm
    \hskip 0.0cm
    \begin{overpic}[width=7.5cm,height=5.0cm,angle=0]{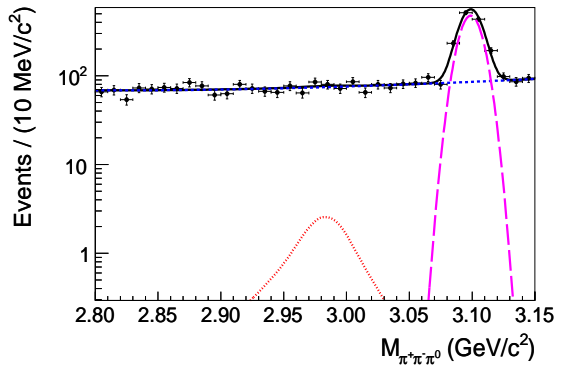}
    \end{overpic}
    {\caption{ The result of the fit on the $\pi^{+}\pi^{-}\pi^{0}$
        mass spectrum in the $\eta_{c}$ region. Dots with error bars
        are data, the solid curve shows the result of unbinned
        maximum likelihood fit, the dotted curve is the $\eta_{c}$
        signal, the long-dashed curve is the $J/\psi$ background, and
        the short-dashed curve is the main background.  }
    \label{etacfit}}
\end{figure}

We perform an unbinned maximum likelihood fit to the
$M_{\pi^{+}\pi^{-}\pi^{0}}$ distribution in the range of [2.80, 3.15]
GeV/$c^{2}$.  In the fit, the $\eta_c$ signal shape is obtained from
exclusive MC samples, the $J/\psi$ background shape is described by a
Breit-Wigner function convolved with a Gaussian function, and the
smooth background is described by a $2^{nd}$-order Chebychev
polynomial function, where all the parameters are free. The fit, shown
in Fig.~\ref{etacfit}, yields $N = 15\pm44$ $\eta_c$-candidate events,
consistent with zero. To obtain an upper limit on the signal
yield, a series of unbinned maximum likelihood fits to the
$M_{\pi^{+}\pi^{-}\pi^{0}}$ distribution are performed for different
values $N$  of the $\eta_c$ signal yield.  The upper limit on $N$ at
the $90\%$ confidence level (C.L.), $N^{UL}_{\eta_c}$, is the value of
$N$ yielding $90\%$ of the integral of the likelihood over all non-negative values
of $N$. The fit-related uncertainties on $N^{UL}_{\eta_c}$ are
considered by varying fit ranges, changing the order of the Chebychev
polynomial function for the background shape and changing the mass and
width of the $\eta_{c}$ within one standard deviation from the central
values for the signal shape. The maximum upper limit amongst the
variations, $N^{UL}_{\eta_{c}}=121$, is used to calculate the upper
limit on the branching fraction.

\subsection{\boldmath \texorpdfstring{$\psi(3686)\to\gamma\eta(1405), \eta(1405)\to f_{0}(980)\pi^{0} $}{eta(1405) to 3pi}}

\begin{figure}[htbp]
    \centering
    \vskip -0.0cm
    \hskip -0.0cm
    \begin{overpic}[width=7.5cm,height=5.0cm,angle=0]{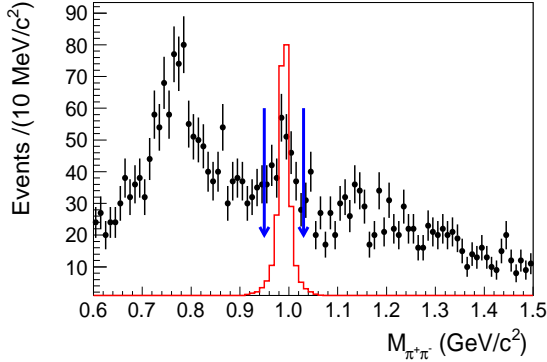}
    \end{overpic}
    {\caption{The $\pi^{+}\pi^{-}$ invariant mass distribution for the
        events with $\pi^{+}\pi^{-}\pi^{0}$ invariant mass within the
        region of [1.20, 2.00] GeV/$c^{2}$. Dots with error bars are
        data, the solid line is the MC $f_0$(980) signal shape, and
        the region between the arrows is the $f_0$(980) mass window. }
    \label{pippim}}
\end{figure}

\begin{figure}[htbp]
    \centering
    \vskip 0.0cm
    \hskip 0.0cm
    \begin{overpic}[width=7.5cm,height=5.0cm,angle=0]{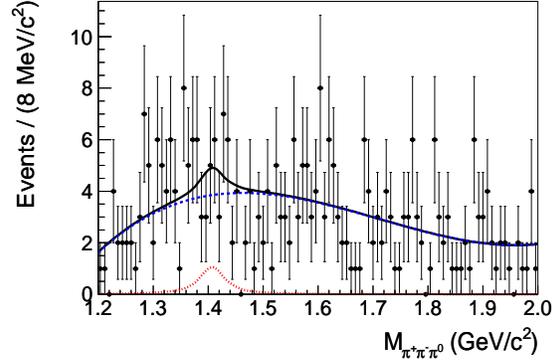}
    \end{overpic}
    {\caption{ Fit to the $\pi^{+}\pi^{-}\pi^{0}$ mass distribution in
        the $\eta(1405)$ region for events satisfying
        $|M_{\pi^{+}\pi^{-}}-m_{f_{0}}|<0.04$ GeV/$c^{2}$. The dots
        with error bars are data, the solid curve shows the result of
        unbinned maximum likelihood fit, the dotted curve is the
        $\eta(1405)$ signal shape, and the short-dashed curve is the
        background. }
    \label{eta1405fit}}
    \end{figure}

The final state for $\psi(3686)\to \gamma \eta(1405)$, $\eta(1405)\to
f_0(980)\pi^0$ with $f_0(980)\to\pi^+\pi^-$ is also $\pi^+ \pi^-
\pi^0$, so we also perform a search for $\eta(1405)\to f_0(980)\pi^0$
in $\psi(3686)$ radiative decays. The same event selection is used for
events with $\pi^{+}\pi^{-}\pi^{0}$ invariant mass within the region
of [1.20, 2.00] GeV/$c^{2}$, and the resulting $\pi^{+}\pi^{-}$
invariant mass distribution is shown in Fig.~\ref{pippim}. A narrow
structure around 0.98 GeV/c$^2$ is observed, which is consistent with
that observed in $J/\psi\to\gamma \eta(1405), \eta(1405)\to
f_0(980)\pi^0$~\cite{wuzhi}.  After requiring the $\pi^+ \pi^-$
invariant mass to satisfy $|M_{\pi^{+}\pi^{-}}-m_{f_{0}}|<0.04$
GeV/$c^{2}$, where $m_{f_{0}}$ is the nominal mass of
$f_0(980)$~\cite{PDG2016}, there is no apparent $\eta(1405)$ signal in
the $M_{\pi^+\pi^-\pi^0}$ distribution, shown in
Fig.~\ref{eta1405fit}.  The background events are investigated using
$\pi^0$ sidebands (0.100 $<M_{\gamma\gamma}<0.115$ GeV/$c^{2}$ and
0.155 $<M_{\gamma\gamma}<0.170$ GeV/$c^{2}$ ), $f_0(980)$ sidebands
(0.90$<M_{\pi^{+}\pi^{-}}<0.94$ GeV/$c^{2}$ and 1.04
$<M_{\pi^{+}\pi^{-}}<1.08$ GeV/$c^{2}$), and the inclusive MC decays,
and no obvious peaking structures are observed around 1.4 GeV/c$^2$.

Using the same approach as in the search for $\eta_c
\to\pi^{+}\pi^{-}\pi^{0} $, we set an upper limit at the 90\% C.L. on
the branching fraction for the decay $\psi(3686) \to \gamma
\eta(1405), \eta(1405)\to f_{0}(980)\pi^{0}$ by fitting the
distribution of $\pi^+\pi^-\pi^0$ invariant mass.  The fit curve is
shown in Fig.~\ref{eta1405fit}, where the signal shape of the
$\eta(1405)$ is obtained from MC simulation in which the mass and
width are fixed to the world average values~\cite{PDG2016}, and the
background is modeled by a $3^{rd}$-order Chebychev polynomial
function. Fit-related uncertainties are determined by performing
various fits with variations of the $\eta(1405)$ mass and width,
different fit ranges and alternative background functions. The largest
upper limit on the yield of $\eta(1405)$ at the $90\%$ C.L. is
$N^{UL}_{\eta(1405)}=38$.

\section{Systematic Uncertainties}
The systematic uncertainties in branching fraction measurements mainly
come from the tracking, photon detection, and particle identification
(PID) efficiencies, the 4C kinematic fit, the $\pi^0$ mass window
requirement, the uncertainties of $\BR(\pi^0\to\gamma\gamma$) and the
number of $\psi(3686)$ events, and the fitting related uncertainties.

The MDC tracking efficiency is studied with clean channels of
$J/\psi\to p\bar{p}\pi^+\pi^-$ and
$J/\psi\to\rho\pi$~\cite{beseta4}, and the MC simulation is found to
agree with data within 1\%. Therefore 2\% is taken as the systematic
uncertainty for the two charged tracks in the final states.

The photon detection efficiency is studied with the control sample
$\jpsi\to\pi^+\pi^-\pi^0$~\cite{photon}. The  difference between data and MC is
less than 1\% per photon. Therefore
3\% is assigned as the systematic uncertainty from the three photons.

The $\pi^\pm$ particle identification efficiency is studied using
a clean control sample of $J/\psi\to\rho\pi$ events, and the PID
efficiency for data agrees with that of the Monte Carlo simulation
within 1\%.  In this analysis, two charged tracks are identified as
pions, so 2\% is taken as the systematic uncertainty.

The uncertainty associated with the 4C kinematic fit comes from the
inconsistency between data and MC simulation; this difference is
reduced by correcting the track helix parameters of the MC simulation,
as described in detail in Ref.~\cite{refsmear}. The correction
parameters for pions are obtained by using control samples of
$\psi(3686)\rightarrow \pi^{+}\pi^{-}\pi^{0}$. In this analysis, the
efficiency obtained from the corrected MC samples is taken as the
nominal value, and we take the differences between the efficiencies
with and without correction, 4.5\% for
$\eta_c\to\pi^+\pi^-\pi^0$, 
and 3.1\% for $\eta(1405)\rightarrow
f_0(980)\pi^0$, as the systematic uncertainties.

The uncertainty due to the width of $f_0(980)$ is estimated by varying
its parameters by 1$\sigma$ in the MC simulation, where the parameters are obtained from the fit to data.
 The relative change of the detection
efficiency, 5.4$\%$, is taken as the corresponding systematic
uncertainty.

The uncertainty related with the $\pi^{0}$ mass window requirement is
studied with control samples of $\psi(3686)\rightarrow
\pi^{+}\pi^{-}\pi^{0}$ for both data and MC simulation.  We fit the
$\gamma\gamma$ invariant mass distribution to determine the $\pi^0$ signal
yields, and the $\pi^0$ efficiency is the ratio of the $\pi^0$ yields
with and without the $\pi^0$ mass window requirement, where the
$\pi^0$ yield is obtained by integrating the fitted signal shape.  The
difference in efficiencies between data and MC simulation, $ 0.8\%$,
is assigned as the systematic uncertainty.

The branching fraction uncertainty of $\pi^{0}\to\gamma\gamma$ is
taken from the PDG~\cite{PDG2016} and is 0.03$\%$. The uncertainty of
the number of $\psi(3686)$ events is 0.65\%~\cite{psi'data}.

For $\eta_c\to\pi^+\pi^-\pi^0$ and $\eta(1405)\rightarrow
f_0(980)\pi^0$, the uncertainties from the fitting range, background
shape, and the signal  shape have already been considered, since we
select the maximum upper limit from amongst various fits described above.

Table ~\ref{summary_of_syserr}  summarizes  all contributions to
the systematic uncertainties on the branching fraction measurements.
 In each case, the total systematic uncertainty is given by the
quadratic sum of the individual contributions, assuming all sources to
be independent.
 \begin {table*}[htp]
    {\caption {Summary of systematic uncertainty sources and their contributions (in \%). 
      }
    \label{summary_of_syserr}}
    \begin{tabular}{l|c|c}  \hline \hline
       Source &  $\eta_{c}\to\pi^{+}\pi^{-}\pi^{0}$ & $\eta(1405)\to f_{0}(980)\pi^{0}$\\ \hline
      MDC tracking  & 2.0 &2.0 \\ \hline
      Photon detection & 3.0 &3.0 \\ \hline
     Particle ID & 2.0  &2.0\\ \hline
      4C kinematic fit & 4.5 & 3.1\\ \hline
      $\pi^{0}$  mass window & 0.8  & 0.8\\ \hline
      Width of $f_{0}(980)$ & ...  & 5.4\\ \hline
      B($\pi^{0}\to\gamma\gamma$) & 0.03 &0.03 \\ \hline
      Number of $\psi(3686)$ & 0.65  &0.65\\ \hline
      Total & 6.2 & 7.5\\ \hline
    \end{tabular}
\end{table*}

\section{Results}

To be conservative,
the upper limit on the branching  fraction is determined  by
\begin{equation}
      \begin{aligned}
      &\mathcal{B}(\psi(3686)\rightarrow \gamma X )\\
      &<\frac{N^{UL}}{N_{\psi(3686)}\times\varepsilon\times \mathcal{B}(\pi^{0}\rightarrow\gamma\gamma)\times(1-\delta_{\rm syst})},
\end{aligned}
\end{equation}
where $X$ stands for $\eta_c(\eta_c\to\pi^+\pi^-\pi^0$) or
$\eta(1405)(\eta(1405)\to f_{0}(980)\pi^{0}\to\pi^{+}\pi^{-}\pi^{0}$),
$\varepsilon$ is the detection efficiency obtained from the MC simulation and $\delta_{\rm syst}$ is the total systematic uncertainty.

The detection efficiencies   are 18.4$\%$ and 18.5$\%$ for $\eta_c \to
\pi^+\pi^-\pi^0$ and $\eta(1405)\to f_0(980)\pi^0$, respectively,
which are determined with MC simulation by assuming the polar angle of
radiative photon follows the distribution $1+\cos^2\theta_\gamma$. The
upper limits at the 90\% C.L. on $\mathcal{B}(\psi(3686)\to \gamma\eta_{c})\times\mathcal{B}(\eta_{c}\to
  \pi^{+}\pi^{-}\pi^{0} )$  and
$\mathcal{B}(\psi(3686)\to\gamma\eta(1405))\times \mathcal{B}(\eta(1405)\to f_{0}(980)\pi^{0})\times \mathcal{B}(f_{0}(980)\to\pi^+\pi^- )$ are calculated to be $1.6\times10^{-6}$
and $5.0 \times10^{-7}$, respectively.

\section{Summary}
Using  448.1$\times$10$^{6}$ $\psi(3686)$ events accumulated with
the BESIII detector, the search for $\eta_{c}\to
\pi^{+}\pi^{-}\pi^{0}$ is performed for the first time. No obvious
$\eta_{c}$ signal is seen in the $\pi^+\pi^-\pi^0$ mass spectrum, and
the 90\% C.L upper limit on $\mathcal{B}(\psi(3686)\to \gamma\eta_{c})\times\mathcal{B}(\eta_{c}\to
  \pi^{+}\pi^{-}\pi^{0} )$  is
$1.6\times10^{-6}$.  Using the branching fraction of
$\psi(3686)\rightarrow \gamma \eta_c$, $[3.4\pm0.5]\times10^{-3}$, the
upper limit for $\mathcal{B}(\eta_c\to\pi^+\pi^-\pi^0)$ is calculated
to be $5.5\times10^{-4}$.  We also search for
$\psi(3686)\to\gamma\eta(1405), \eta(1405)\to f_0(980)\pi^0$. No
obvious structure around the $\eta(1405)$ is observed, and the 90\% C.L
 upper limit on $\mathcal{B}(\psi(3686)\to\gamma\eta(1405))\times \mathcal{B}(\eta(1405)\to f_{0}(980)\pi^{0})\times \mathcal{B}(f_{0}(980)\to\pi^+\pi^- )$ is $5.0 \times10^{-7}$.  In addition, based on the measurement in $J/\psi$
decays~\cite{wuzhi}, the ratio of
$\frac{\mathcal{B}(\psi(3686)\to\gamma\eta(1405))}
{\mathcal{B}(J/\psi\to\gamma\eta(1405))}$ is calculated to be less
than $3.3\times 10^{-2}$, which indicates that this process also violates the ``12\% rule''.

\begin{acknowledgments}
The BESIII collaboration thanks the staff of BEPCII and the IHEP computing center for their strong support. This work is supported in part by National Key Basic Research Program of China under Contract No. 2015CB856700; National Natural Science Foundation of China (NSFC) under Contracts Nos. 11735014, 11675184, 11235011, 11322544, 11335008, 11425524, 11635010; the Chinese Academy of Sciences (CAS) Large-Scale Scientific Facility Program; the CAS Center for Excellence in Particle Physics (CCEPP); the Collaborative Innovation Center for Particles and Interactions (CICPI); Joint Large-Scale Scientific Facility Funds of the NSFC and CAS under Contracts Nos. U1232201, U1332201, U1532257, U1532258; CAS under Contracts Nos. KJCX2-YW-N29, KJCX2-YW-N45, QYZDJ-SSW-SLH003; 100 Talents Program of CAS; National 1000 Talents Program of China; INPAC and Shanghai Key Laboratory for Particle Physics and Cosmology; German Research Foundation DFG under Contracts Nos. Collaborative Research Center CRC 1044, FOR 2359; Istituto Nazionale di Fisica Nucleare, Italy; Koninklijke Nederlandse Akademie van Wetenschappen (KNAW) under Contract No. 530-4CDP03; Ministry of Development of Turkey under Contract No. DPT2006K-120470; National Science and Technology fund; The Swedish Resarch Council; U. S. Department of Energy under Contracts Nos. DE-FG02-05ER41374, DE-SC-0010118, DE-SC-0010504, DE-SC-0012069; University of Groningen (RuG) and the Helmholtzzentrum fuer Schwerionenforschung GmbH (GSI), Darmstadt; WCU Program of National Research Foundation of Korea under Contract No. R32-2008-000-10155-0.
\end{acknowledgments}


\end{document}